\documentclass[twocolumn,showpacs,preprintnumbers,amsmath,amssymb]{revtex4}

\usepackage{graphicx}
\usepackage{dcolumn}
\usepackage{bm}

\begin{document}


\title{On the phase of magneto-oscillations in graphite}
\author{L.\ Smr\v{c}ka}
\affiliation{Institute of Physics, Academy of Science of the Czech
Republic,\nolinebreak[5] v.v.i.,\\ 
Cukrovarnick\'{a} 10, 162 53 Prague 6, Czech Republic}
\author{N.\ A.\ Goncharuk}
\affiliation{Institute of Physics, Academy of Science of the Czech
Republic,\nolinebreak[5] v.v.i.,\\ 
Cukrovarnick\'{a} 10, 162 53 Prague 6, Czech Republic}

\date{\today}

\begin{abstract}
The problem of Dirac fermions in graphite subject to a perpendicular
magnetic field is studied. We show analytically that the weak
inter-layer interaction between the graphene sheets leads to anomalies
in the Shubnikov-de Haas and de Haas-van Alphen magneto-oscillations
governed by the orbits around extremal cross-sections of the graphite
Fermi surface. The calculation of the Landau plot performed within a
four band continuum model reveals that magneto-oscillations are
aperiodic, except of the case of vanishing inter-layer interaction at
the H point of the graphite Brillouin zone. Also for all other orbits
along the H-K-H edge the magneto-oscillations are only
asymptotically periodic in the quasi-classical limit, with the phase
corresponding to massive fermions.
\end{abstract}

\maketitle

%
Graphite is a layered material composed of weakly coupled
two-dimensional (2D) graphene sheets formed by hexagonally arranged
carbon atoms. In 2004, a single sheet of graphene was prepared from
three-dimensional (3D) graphite by micro-mechanical cleavage
\cite{geim04}. The discovery immediately attracted attention of the
solid-state physical community, as the electrons in graphene obey a
linear energy dependence on the wave-vector $k$, and behave like
massless relativistic particles -- Dirac fermions (DFs). In the
seminal papers \cite{geim04, geim05,kim05}, the Shubnikov-de Haas
(SdH) magneto-oscillations in graphene were found periodic in an
inverse magnetic field, $1/B$, similarly as in the case of a 2D gas of
massive Schr\"{o}dinger fermions (SFs), but with the phase shifted by
$\pi$. The shift, which was clearly demonstrated by the Landau plot of
magneto-resistance oscillations, is due to the existence of the
zero-energy Landau level (LL), shared by electrons and holes.  For the
same reason, the anomalous quantum Hall effect with a half-integer
instead of integer quantization was observed in mechanically
ex-foliated samples \cite{geim05}. This is considered as the most
direct evidence of DFs in graphene.  In 2006,  important
technological progress was achieved. The epitaxial graphene was grown
on the single-crystal silicon carbide by vacuum graphitization
\cite{berger06, berger07}.
 
The discovery of DFs in graphene has resulted in renewed interest in
bulk graphite.  In a series of papers \cite{luk04, luk06, luk09}, the
spectral analysis of SdH and de Haas-van Alphen (dHvA) oscillations
was employed to determine the phases of two series of
magneto-oscillations observed in graphite. In papers
\cite{miki99,miki06} an attempt was made to relate the phases to the
topological Berry phase, which is acquired by fermions moving around
close orbits.  Based on their analysis the authors of Refs.~\cite{luk04,
luk06, luk09} came to a conclusion that one of two groups of
oscillating carriers corresponds to DFs.  Recently a paper \cite{sch}
was devoted to a careful, mostly experimental investigation of SdH
effect in graphite, and doubts about the observation of DFs using
magneto-transport measurements were expressed.

Here the problem is treated from the theoretical point of view. We
construct a Landau plot for the  model Hamiltonian developed by
Slonczewski, Weiss and MacClure (SWM) \cite{slon,mcclure57,mcclure60}
and compare the result with Landau plots for SFs and DFs,
described below.
 
%
The energy spectra of 2D SFs and DFs in a zero magnetic field can be
written as
\begin{equation} 
E^{S}(k) = \frac{\hbar^2k^2}{2m^*}, \,\,\,\ E^{D}(k) =\pm \hbar v_F k,
\label{zeroB}
\end{equation}
where $k=\sqrt{k_x^2+k_y^2}$, $m^*$ is the effective mass of SFs and
$v_F$ is the Fermi velocity of DFs. The positive and negative branches
of the conical DFs spectrum correspond to the electrons and holes,
respectively.
 
In a magnetic field, the spectra of SFs and DFs are quantized into the
LLs as follows:
\begin{equation} 
E^{S}_{n} = \hbar\omega_c\left(n+\frac{1}{2}\right), \,\,\,\ 
E^{D}_{n} =\pm \sqrt{2\hbar|e| v_F^2 B n},
\label{nonzeroB}
\end{equation}
where $\omega_c = |e|B/m^*$ is the cyclotron frequency and the index
$n=0,1,2,\cdots$.  In the case of SFs the equidistant LLs lie above
$E=0$ for any finite $B$, whereas in the DFs case the lowest electron
LL is shared with the highest hole LL located exactly at $E=0$.

The Eqs.~(\ref{nonzeroB}) are consistent with the Onsager-Lifshitz
quasi-classical quantization rule 
\begin{equation} 
A^Q(E_F)= \frac{2\pi|e|B}{\hbar}\left(n+\gamma^Q\right), \,\,\,
Q =S,D,
\label{onsager}
\end{equation}
where $A^Q(E_F)=\pi k_F^2$ is the area of the SF or DF Fermi circle,
calculated with the Fermi energy $E_F$ and the Fermi wave-vector $k_F$
taken from Eqs.~(\ref{zeroB}). We get $\gamma^S=1/2$ for SFs and 
$\gamma^D=0$ for DFs.

Magneto-oscillations observed in SdH and dHvA effects are controlled
by oscillations of the density of states (DOS).  It is well
known that the DOS on the Fermi level, $g(E_F)$, can be expressed as an
imaginary part of the resolvent $G(z)=(z-H)^{-1}$,
\begin{equation} 
g(E_F)= -\frac{1}{\pi}\, {\mathcal Im}\, {\text Tr}\,G(E_F+i0 ).
\label{res}
\end{equation}
For simple diagonal  Hamiltonians of SFs and DFs given by
Eqs.~(\ref{nonzeroB}), we get
\begin{equation}
g^Q(E_F)=\frac{|e|B}{2\pi\hbar}\,\sum_{n=0}^{\infty} 
\delta\left(E_F -E^Q_n\right).
\label{doss}
\end{equation}
It follows from Eq.~(\ref{doss}) that the DOS reaches maxima at
magnetic fields $B_n$ for which the LLs cross the Fermi energy $E_F$.
A Landau plot, i.e., the plot of the inverse magnetic fields $1/B_n$
versus the level index $n$ is a standard tool used to determine the
frequency and phase of magneto-oscillations. For SFs and DFs we arrive
to
\begin{equation}
\frac{B^{Q}_{0}}{B_n} = n+\gamma^Q, 
\label{period}
\end{equation}
where $B^{S}_{0} =m^{*} E_F /(\hbar |e|)$ and $B^{D}_{0}=E_F^2/(2\hbar
|e|v_F^2)$ are the oscillation frequencies, in agreement with the
quasi-classical expression obtained from Eq.~(\ref{onsager}),
\begin{equation} 
B^{Q}_{0}= \hbar A^Q/2\pi|e|.
\label{quasi}
\end{equation} 
It is clear that the positions of maxima of SF and DF oscillations
differ by a half of the period, i.e., by $\pi$ in terms of a phase
factor.

In graphite, the inter-layer interaction of Bernald-stacked graphenes
adds a $k_z$-dependence to the electron energy spectrum and a 3D Fermi
surface (FS) is formed close to the H-K-H edge of the hexagonal
Brillouin zone (BZ).  As mentioned above, the graphite 3D electronic
structure is described by the semi-empirical SWM Hamiltonian, which
employs seven nearest-neighbor tight-binding (TB) parameters
$\gamma_0,\gamma_1,\dots\gamma_5, \Delta$, and the value of the Fermi
energy, $E_F$. Previously, the model parameters were fitted to various
optical and transport experiments \cite{dress81}. Recently, their
values are continuously refined by fitting to the experimental data
\cite {sch} and/or to the results of first-principles numerical
simulations of the graphite band structure \cite{gruneis}.

Among the seven SWM parameters, the parameter $\gamma_3$, which
controls the trigonal warping of the FS, brings a numerical
complications in the case of nonzero magnetic field. When $\gamma_3$
is taken into account, the magnetic-field-SWM Hamiltonian has an
infinite order and must be diagonalized numerically
\cite{nakao}. Fortunately, its influence is not too strong for $k_z$
far from the H point of the BZ and energies close to $E_F$ \cite{nakao}.  
To facilitate our analytical treatment, we prefer to use a
simplified Hamiltonian $H$ with $\gamma_3$ neglected, in the form
introduced in the MacClure's paper \cite{mcclure60}.

The choice of $\gamma_3 = 0$ yields isotropic equienergetic
contours. The FS of graphite consists of elongated electron and
hole pockets located near the points K and H, with $k_z$-dependent
circular cross-sections.  In the quasi-classical limit, two extremal
cross-sections define two series of magneto-oscillations and two
quasi-classical frequencies $B_0^e$ and $B_0^h$ for electrons and holes,
respectively.  Analytical solutions for the FS cross-sections can be
found, e.g., in Ref.~\cite{dress81}.

Within these approximations an expression similar to
Eq.~(\ref{period}) can be obtained for any $k_z$-dependent
cross-section of the 3D graphite FS. To do so, we need to find the
poles of the resolvent $G(z)=(E_F-H+i0)^{-1}$. In other words, we
should solve the secular equation derived from the simplified
Hamiltonian $H$ of Ref.~\cite{mcclure60} for $B_n$, i.e., we should
find the roots $B_n$ of the secular polynomial.
\begin{figure}[t]
\includegraphics[width=\linewidth]{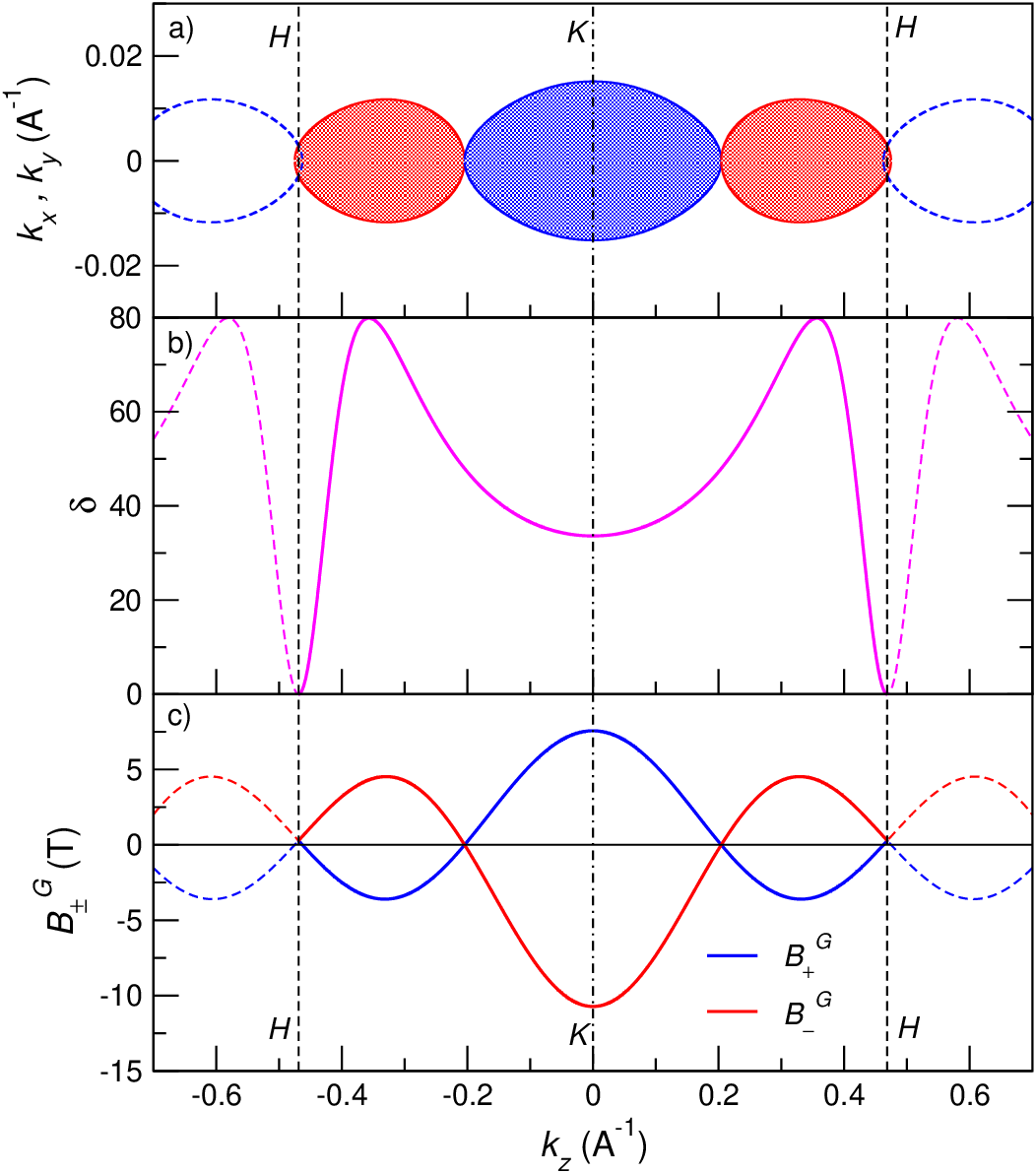}
\caption{\label{delta}(color online) a) Contours of the graphite Fermi
surface, b) a dimensionless parameter $\delta$ as a function of
$k_z$, c) the parameters $B^{G}_{\pm}$ as functions of $k_z$. }
\end{figure}

The solution  yields a formula for the inverse magnetic fields as a
function of the level index, $n$, in a shape
\begin{equation} 
\frac{B^{G}_{\pm}(k_z)}{B_n} = \frac{
n+\frac{1}{2}\pm\sqrt{\frac{1}{4}+n(n+1)\delta(k_z)}
}{1\pm\sqrt{\delta(k_z)}},
\label{aperiod}
\end{equation}
where the three coefficients $B^{G}_{\pm}(k_z)$ and $\delta(k_z)$ can be
derived from the SWM model and the value of $E_F$. The $k_z$-dependence
originates from $\cos{(k_zc/2)}$ which appears in the inter-layer TB SWM
parameters, $c/2$ denotes the inter-layer distance in
graphite. Obviously, $B^{G}_{\pm}(k_z)/B_n$, as given by
Eq.~(\ref{aperiod}), are not periodic in $1/B$.

\begin{figure}[htb]
\includegraphics[width=\linewidth]{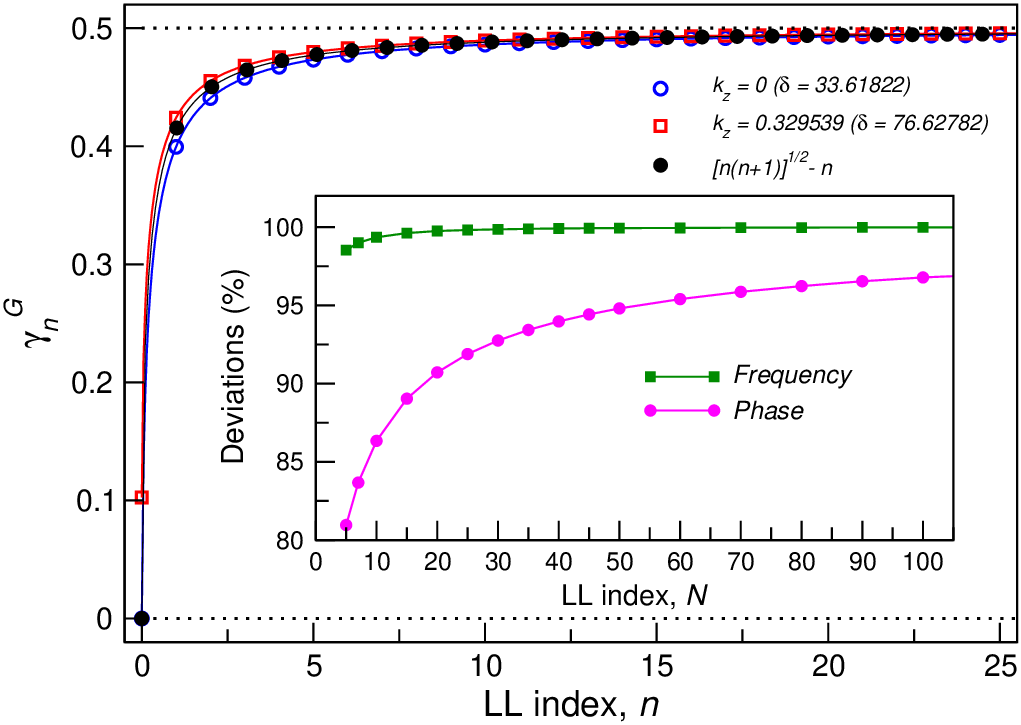}
\caption{\label{gamma}(color online) $\gamma^G_{\pm,n}(k_z)$ for maximum
cross-sections of the electron and hole pockets approximated by
$\sqrt{n(n+1)}-n$.  The inset shows results of the linear
approximation of first $N$ terms ($n=1,2,\cdots,N$) of
$\sqrt{n(n+1)}$. The relative deviations from the quasi-classical
frequency and phase are shown.}
\end{figure}

An expression for the Landau plot corresponding to Eq.~(\ref{aperiod})
can be written in a form
\begin{equation} 
\frac{B^{G}_{\pm}(k_z)}{B_n} =n+\gamma^G_{\pm,n}(k_z).
\label{landau}
\end{equation}
In this equation $\gamma^G_{\pm,n}(k_z)$, defined by the right-hand
side of Eq.~(\ref{aperiod}), is no longer a constant describing the
oscillation phase, but a variable which depends on the LL index $n$.

The dependence on $n$ is most pronounced for high magnetic fields,
i.e., for small $n$.  In the low-magnetic-field limit, with a large
number of LLs below $E_F$, we can write $n(n+1)\delta(k_z)\gg 1/4$ and
$\sqrt{n(n+1)}\rightarrow n+1/2$.  Then $\gamma^G_{\pm,n}(k_z) \rightarrow
1/2$, and we can conclude that the charge carriers in graphite behave,
at least as far as  the phase is concerned, as the SFs.

Only when we can completely neglect the inter-layer interaction, as at
the $H$ point of the 3D BZ, $k_z = \pi/c$, $\cos{(k_zc/2)}=0$,
$\Delta=0$ and $\delta(k_z) \rightarrow 0$, we get
\begin{equation} 
\frac{B^{G}_{\pm}(\pi/c)}{B_n} = 
n+\frac{1}{2}\pm\frac{1}{2},
\label{Hpoint}
\end{equation} 
a result which corresponds to DFs.

Here the Landau plots of two series of magneto-oscillations,
which can be observed in graphite, are considered as most
interesting. We constructed them based on the parameters of the SWM
model taken from Ref.~\cite{dress81}.  The explicit expressions for
$B^{G}_{\pm}$ and $\delta(k_z)$, which appear in Eq.~(\ref{aperiod}),
will not be presented here. Instead, their $k_z$-dependences are shown
in Fig.~\ref{delta}, together with the contours of the FS.  The
positive sign applies in the formula (\ref{aperiod}) for $k_z$ from
the electron region of FS and the parameter $B^{G}_{+}$ is equal to
the quasi-classical frequency $B_0^e$.  Similarly, the negative sign
should be taken for $k_z$ from the hole pockets, where $B^{G}_{-}$
becomes equal to $B_0^h$.  For both electron and hole extremal orbits
the parameter $\delta$ is large and, consequently, the right-hand side
of Eq.~(\ref{aperiod}) can be approximated by $\sqrt{n(n+1)}$. The
marked difference is only for $n=0$. The Fig.~\ref{gamma} shows to
what extent the accuracy of this approximation is reasonable.

The expression for $B^{G}_{\pm}/B_n$ is not periodic and the question
arises how many LLs must be resolved to reach the linear dependence on
$n$, i.e., the quasi-classical limit $n+1/2$.  The result of the
linear approximation of a model curve $\sqrt{n(n+1)}$ is presented in
the inset of Fig.~\ref{gamma}, where the relative deviations from the
quasi-classical frequency and phase are shown. It turns out that both
frequency and phase are underestimated if we took into account only
limited number of $n$, $1\le n \le N$.  Less oscillations are
necessary to get close to the quasi-classical frequency than to obtain
a reasonable approximation for the phase.  This may explain the
differences found between the experimentally determined phases of
samples with different mobilities, which are determined from different
number of oscillations resolved.

Note that the above Landau plot can be derived if we approximate the
energy spectra of electrons and holes by the formulae
\begin{equation} 
E^e_n = \hbar \omega^e_c\sqrt{n(n+1)},\,\,\,
E^h_n = \hbar \omega^h_c\sqrt{n(n+1)}, 
\label{elhole}
\end{equation} 
where $\omega^e_c$ and $\omega^h_c$ are the quasi-classical cyclotron 
frequencies corresponding to extremal electron and hole orbits.
\begin{figure}[b]
\includegraphics[width=\linewidth]{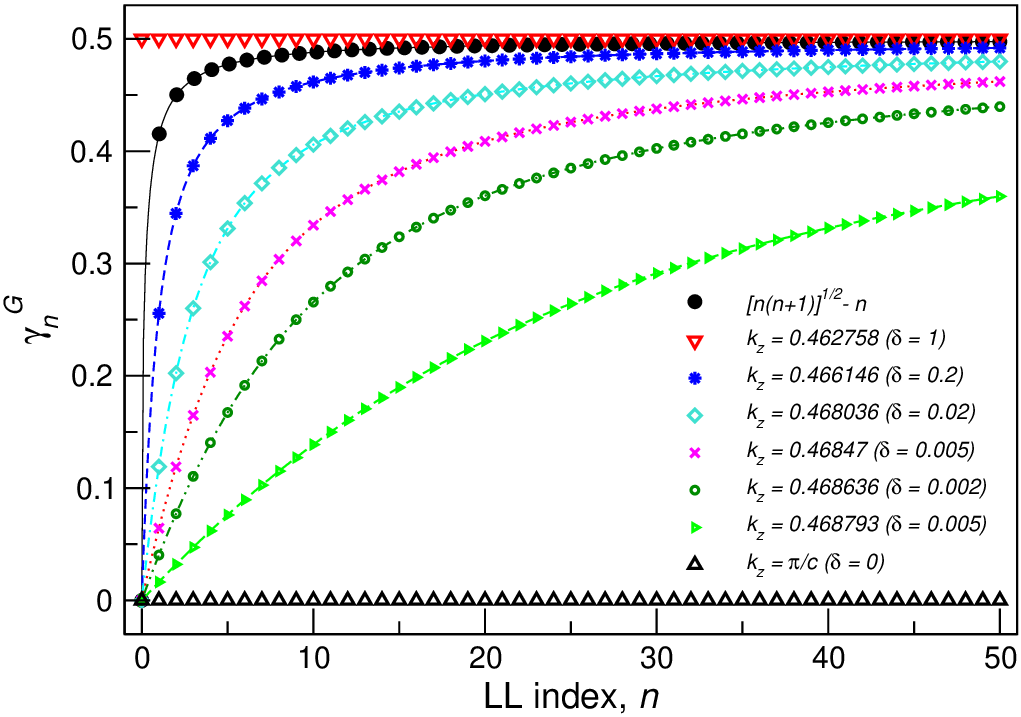}
\caption{\label{gammaH}(color online)The dependence of
$\gamma_{-,n}(k_z)$ on the LL index $n$ for $k_z$ close to the H point
of the BZ. }
\end{figure}

Two series of magneto-oscillations discussed above correspond to
$\delta \gg 1$ and to the maximum cross-sections of the electron and
hole pockets. There are another two extremal cross-sections, till now
not reliably resolved in the transport experiments, located around the
H point of the BZ where $\delta \le 1$. In spite of the fact that,
according to Ref.~\cite{nakao}, the parameter $\gamma_3$ has a
qualitative influence on the LLs structure near this point, it is at
least of the theoretical interest to study  the behavior of the
$\gamma^G_{\pm,n}(k_z)$  for  model with $\gamma_3$ neglected.

For a given $k_z$ the dependence of the energy bands on $k$ is
hyperbolic in a zero magnetic field. While near the maximum
cross-sections there are broad minima/maxima of bands which are
similar to parabolas for small $k$, near the H point the hyperbolas
are very sharp and with a shape close to the Dirac cone. Therefore, a
smooth transition of $\gamma^G_{\pm,n}(k_z)$ from SFs to DFs is
expected. Fig.~\ref{gammaH} reveals that the behavior is more
complicated.

In the SWM model the $k$-dependence of the zero-field  energy at the H point
($k_z=0$) is given by
\begin{equation} 
E^G = \frac{\Delta}{2} \pm \sqrt{\left(\frac{\Delta}{2}\right)^2 
+\hbar^2 v_F^2k^2},
\label{kz0}
\end{equation}
which is not equal to the Dirac cone for a finite $\Delta$.
Nevertheless, $\delta =0$ implies that $\gamma^G_{-,n}(k_z)$ is a
constant equal to $0$, as for the DFs. This is in agreement with the
Landau plot constructed from the energy spectra in a magnetic field,
which according to Ref.~\cite{mcclure60} have a simple analytic form
\begin{equation} 
E_n^G = \frac{\Delta}{2} \pm \sqrt{\left(\frac{\Delta}{2}\right)^2 
+2\hbar|e| v_F^2Bn}.
\label{kz0B}
\end{equation}

On the other hand, the parameter $\delta$ equals 1 for $E_F$ crossing
the $E_2$ band of the SWM model. According to Eq.~(\ref{aperiod}) this
leads to $\gamma^G_{-,n}(k_z) = 1/2$, as for the SFs, in spite of the
$k$-dependence not so close to parabolic one as the for the maximum
cross-sections, $\delta \gg 1$, where the energy spectra can be
approximated by Eqs.~(\ref{elhole}).

The field dependence of the corresponding LLs ranges from that
described by Eq.~(\ref{kz0B}) at the H point, which is close to
$\sqrt{B}$ characteristic for DFs, to the linear dependence on $B$
typical for SFs for extremal electron and hole orbits, as given by
Eqs.~(\ref{elhole}).

It follows from the above discussion that the hyperbolic
$k$-dependence of the zero-field electron energy bands, which changes
considerably depending on the value of $k_z$, yields aperiodic
magneto-oscillations when a magnetic field is applied. An exception
are two $k_z$ in the neighborhood of the H point of the BZ. We assume
that this conclusion is at least qualitatively correct, as one can
hardly believe that this is just the neglected $\gamma_3$ which yields
the magneto-oscillation aperiodicity.

There is another potential reason for deviations from the
magneto-oscillation periodicity.  Unlike the optical experiments which
involve electrons with energies below and above the Fermi energy, the
SdH and dHvA magneto-oscillations reflect only the properties of
electrons with an energy equal to $E_F$.  Our treatment is
based on the assumption that the $E_F$ is a constant.  This is not
quite correct as the carrier concentration is a constant and not
$E_F$, which should oscillate as a function of $B$. This can be
important for lowest LLs in high mobility samples and was considered
as a single source of oscillation aperiodicity in Ref.~\cite{sch}.

In conclusion, we have found that the magneto-oscillations in graphite are
only asymptotically periodic in the quasi-classical limit, with the
phase corresponding to massive fermions. The quasi-classical limit can
be reached only exceptionally for samples with very high mobility and
at very low magnetic field. Therefore, the determination of the
oscillation phase in samples with a limited number of resolved
LLs below $E_F$ is not a reliable tool for distinguishing between DFs
and SFs in graphite, due to the aperiodicity of the magneto-oscillations
in a standard quantum regime.
 
The authors acknowledge the support of the Academy of Sciences of the
Czech Republic project KAN400100652 and the Ministry of Education of
the Czech Republic project LC510.

\bibliography{text.bib}

\begin{thebibliography}{17}
\expandafter\ifx\csname natexlab\endcsname\relax\def\natexlab#1{#1}\fi
\expandafter\ifx\csname bibnamefont\endcsname\relax
  \def\bibnamefont#1{#1}\fi
\expandafter\ifx\csname bibfnamefont\endcsname\relax
  \def\bibfnamefont#1{#1}\fi
\expandafter\ifx\csname citenamefont\endcsname\relax
  \def\citenamefont#1{#1}\fi
\expandafter\ifx\csname url\endcsname\relax
  \def\url#1{\texttt{#1}}\fi
\expandafter\ifx\csname urlprefix\endcsname\relax\def\urlprefix{URL }\fi
\providecommand{\bibinfo}[2]{#2}
\providecommand{\eprint}[2][]{\url{#2}}

\bibitem[{\citenamefont{Novoselov et~al.}(2004)\citenamefont{Novoselov, Geim,
  Morozov, D.~Jiang, Dubonos, Grigorieva, and Firsov}}]{geim04}
\bibinfo{author}{\bibfnamefont{K.~S.} \bibnamefont{Novoselov}},
  \bibinfo{author}{\bibfnamefont{A.~K.} \bibnamefont{Geim}},
  \bibinfo{author}{\bibfnamefont{S.~V.} \bibnamefont{Morozov}},
  \bibinfo{author}{\bibfnamefont{Y.~Z.} \bibnamefont{D.~Jiang}},
  \bibinfo{author}{\bibfnamefont{S.~V.} \bibnamefont{Dubonos}},
  \bibinfo{author}{\bibfnamefont{I.~V.} \bibnamefont{Grigorieva}},
  \bibnamefont{and} \bibinfo{author}{\bibfnamefont{A.~A.}
  \bibnamefont{Firsov}}, \bibinfo{journal}{Science}
  \textbf{\bibinfo{volume}{306}}, \bibinfo{pages}{666} (\bibinfo{year}{2004}).

\bibitem[{\citenamefont{Novoselov et~al.}(2005)\citenamefont{Novoselov, Geim,
  Morozov, Jiang, Katsnelson, Grigorieva, Dubonos, and Firsov}}]{geim05}
\bibinfo{author}{\bibfnamefont{K.~S.} \bibnamefont{Novoselov}},
  \bibinfo{author}{\bibfnamefont{A.~K.} \bibnamefont{Geim}},
  \bibinfo{author}{\bibfnamefont{S.~V.} \bibnamefont{Morozov}},
  \bibinfo{author}{\bibfnamefont{D.}~\bibnamefont{Jiang}},
  \bibinfo{author}{\bibfnamefont{M.~I.} \bibnamefont{Katsnelson}},
  \bibinfo{author}{\bibfnamefont{I.~V.} \bibnamefont{Grigorieva}},
  \bibinfo{author}{\bibfnamefont{S.~V.} \bibnamefont{Dubonos}},
  \bibnamefont{and} \bibinfo{author}{\bibfnamefont{A.~A.}
  \bibnamefont{Firsov}}, \bibinfo{journal}{Nature}
  \textbf{\bibinfo{volume}{438}}, \bibinfo{pages}{197} (\bibinfo{year}{2005}).

\bibitem[{\citenamefont{Zhang et~al.}(2005)\citenamefont{Zhang, Tan, Stormer,
  and Kim}}]{kim05}
\bibinfo{author}{\bibfnamefont{Y.}~\bibnamefont{Zhang}},
  \bibinfo{author}{\bibfnamefont{Y.-W.} \bibnamefont{Tan}},
  \bibinfo{author}{\bibfnamefont{H.~L.} \bibnamefont{Stormer}},
  \bibnamefont{and} \bibinfo{author}{\bibfnamefont{P.}~\bibnamefont{Kim}},
  \bibinfo{journal}{Nature} \textbf{\bibinfo{volume}{438}},
  \bibinfo{pages}{201} (\bibinfo{year}{2005}).

\bibitem[{\citenamefont{Berger et~al.}(2006)\citenamefont{Berger, Song, Li, Wu,
  Brown, Naud, Mayou, Li, Haas, Marchenkov et~al.}}]{berger06}
\bibinfo{author}{\bibfnamefont{C.}~\bibnamefont{Berger}},
  \bibinfo{author}{\bibfnamefont{Z.}~\bibnamefont{Song}},
  \bibinfo{author}{\bibfnamefont{X.}~\bibnamefont{Li}},
  \bibinfo{author}{\bibfnamefont{X.}~\bibnamefont{Wu}},
  \bibinfo{author}{\bibfnamefont{N.}~\bibnamefont{Brown}},
  \bibinfo{author}{\bibfnamefont{C.}~\bibnamefont{Naud}},
  \bibinfo{author}{\bibfnamefont{D.}~\bibnamefont{Mayou}},
  \bibinfo{author}{\bibfnamefont{T.}~\bibnamefont{Li}},
  \bibinfo{author}{\bibfnamefont{J.}~\bibnamefont{Haas}},
  \bibinfo{author}{\bibfnamefont{A.~N.} \bibnamefont{Marchenkov}},
  \bibnamefont{et~al.}, \bibinfo{journal}{Science}
  \textbf{\bibinfo{volume}{312}}, \bibinfo{pages}{1191} (\bibinfo{year}{2006}).

\bibitem[{\citenamefont{Berger et~al.}(2007)\citenamefont{Berger, Song, Li, Wu,
  Brown, Maude, Naud, and de~Heer}}]{berger07}
\bibinfo{author}{\bibfnamefont{C.}~\bibnamefont{Berger}},
  \bibinfo{author}{\bibfnamefont{Z.}~\bibnamefont{Song}},
  \bibinfo{author}{\bibfnamefont{X.}~\bibnamefont{Li}},
  \bibinfo{author}{\bibfnamefont{X.}~\bibnamefont{Wu}},
  \bibinfo{author}{\bibfnamefont{N.}~\bibnamefont{Brown}},
  \bibinfo{author}{\bibfnamefont{D.~K.} \bibnamefont{Maude}},
  \bibinfo{author}{\bibfnamefont{C.}~\bibnamefont{Naud}}, \bibnamefont{and}
  \bibinfo{author}{\bibfnamefont{W.~A.} \bibnamefont{de~Heer}},
  \bibinfo{journal}{phys.\ stat.\ sol. (a)} \textbf{\bibinfo{volume}{204}},
  \bibinfo{pages}{1746} (\bibinfo{year}{2007}).

\bibitem[{\citenamefont{Luky'anchuk and Kopelevich}(2004)}]{luk04}
\bibinfo{author}{\bibfnamefont{I.~A.} \bibnamefont{Luky'anchuk}}
  \bibnamefont{and}
  \bibinfo{author}{\bibfnamefont{Y.}~\bibnamefont{Kopelevich}},
  \bibinfo{journal}{Phys.\ Rev.\ Lett.} \textbf{\bibinfo{volume}{93}},
  \bibinfo{pages}{166402} (\bibinfo{year}{2004}).

\bibitem[{\citenamefont{Luky'anchuk and Kopelevich}(2006)}]{luk06}
\bibinfo{author}{\bibfnamefont{I.~A.} \bibnamefont{Luky'anchuk}}
  \bibnamefont{and}
  \bibinfo{author}{\bibfnamefont{Y.}~\bibnamefont{Kopelevich}},
  \bibinfo{journal}{Phys.\ Rev.\ Lett.} \textbf{\bibinfo{volume}{97}},
  \bibinfo{pages}{256801} (\bibinfo{year}{2006}).

\bibitem[{\citenamefont{Luky'anchuk et~al.}(2009)\citenamefont{Luky'anchuk,
  Kopelevich, and Marssi}}]{luk09}
\bibinfo{author}{\bibfnamefont{I.~A.} \bibnamefont{Luky'anchuk}},
  \bibinfo{author}{\bibfnamefont{Y.}~\bibnamefont{Kopelevich}},
  \bibnamefont{and} \bibinfo{author}{\bibfnamefont{M.~E.}
  \bibnamefont{Marssi}}, \bibinfo{journal}{Physica B}
  \textbf{\bibinfo{volume}{404}}, \bibinfo{pages}{404} (\bibinfo{year}{2009}).

\bibitem[{\citenamefont{Mikitik and Sharlai}(1999)}]{miki99}
\bibinfo{author}{\bibfnamefont{G.~P.} \bibnamefont{Mikitik}} \bibnamefont{and}
  \bibinfo{author}{\bibfnamefont{Y.~V.} \bibnamefont{Sharlai}},
  \bibinfo{journal}{Phys.\ Rev.\ Lett.} \textbf{\bibinfo{volume}{82}},
  \bibinfo{pages}{2147} (\bibinfo{year}{1999}).

\bibitem[{\citenamefont{Mikitik and Sharlai}(2006)}]{miki06}
\bibinfo{author}{\bibfnamefont{G.~P.} \bibnamefont{Mikitik}} \bibnamefont{and}
  \bibinfo{author}{\bibfnamefont{Y.~V.} \bibnamefont{Sharlai}},
  \bibinfo{journal}{Phys.\ Rev.\ B} \textbf{\bibinfo{volume}{73}},
  \bibinfo{pages}{235112} (\bibinfo{year}{2006}).

\bibitem[{\citenamefont{Schneider et~al.}()\citenamefont{Schneider, Orlita,
  Potemski, and Maude}}]{sch}
\bibinfo{author}{\bibfnamefont{J.~M.} \bibnamefont{Schneider}},
  \bibinfo{author}{\bibfnamefont{M.}~\bibnamefont{Orlita}},
  \bibinfo{author}{\bibfnamefont{M.}~\bibnamefont{Potemski}}, \bibnamefont{and}
  \bibinfo{author}{\bibfnamefont{D.~K.} \bibnamefont{Maude}},
  \eprint{arXiv:0902.1925}.

\bibitem[{\citenamefont{Slonczewski and Weiss}(1958)}]{slon}
\bibinfo{author}{\bibfnamefont{C.}~\bibnamefont{Slonczewski}} \bibnamefont{and}
  \bibinfo{author}{\bibfnamefont{P.~R.} \bibnamefont{Weiss}},
  \bibinfo{journal}{Phys.\ Rev.} \textbf{\bibinfo{volume}{109}},
  \bibinfo{pages}{272} (\bibinfo{year}{1958}).

\bibitem[{\citenamefont{McClure}(1957)}]{mcclure57}
\bibinfo{author}{\bibfnamefont{J.~W.} \bibnamefont{McClure}},
  \bibinfo{journal}{Phys.\ Rev.} \textbf{\bibinfo{volume}{108}},
  \bibinfo{pages}{612} (\bibinfo{year}{1957}).

\bibitem[{\citenamefont{McClure}(1960)}]{mcclure60}
\bibinfo{author}{\bibfnamefont{J.~W.} \bibnamefont{McClure}},
  \bibinfo{journal}{Phys.\ Rev.} \textbf{\bibinfo{volume}{119}},
  \bibinfo{pages}{606} (\bibinfo{year}{1960}).

\bibitem[{\citenamefont{Dresselhaus and Dresselhaus}(1981)}]{dress81}
\bibinfo{author}{\bibfnamefont{M.~S.} \bibnamefont{Dresselhaus}}
  \bibnamefont{and}
  \bibinfo{author}{\bibfnamefont{G.}~\bibnamefont{Dresselhaus}},
  \bibinfo{journal}{Advances in Phys. 30} \textbf{\bibinfo{volume}{30}},
  \bibinfo{pages}{139} (\bibinfo{year}{1981}).

\bibitem[{\citenamefont{Gruneis et~al.}(2008)\citenamefont{Gruneis,
  Attaccalite, Wirtz, Shiozawa, Saito, Pichler, and Rubio}}]{gruneis}
\bibinfo{author}{\bibfnamefont{A.}~\bibnamefont{Gruneis}},
  \bibinfo{author}{\bibfnamefont{C.}~\bibnamefont{Attaccalite}},
  \bibinfo{author}{\bibfnamefont{L.}~\bibnamefont{Wirtz}},
  \bibinfo{author}{\bibfnamefont{H.}~\bibnamefont{Shiozawa}},
  \bibinfo{author}{\bibfnamefont{R.}~\bibnamefont{Saito}},
  \bibinfo{author}{\bibfnamefont{T.}~\bibnamefont{Pichler}}, \bibnamefont{and}
  \bibinfo{author}{\bibfnamefont{A.}~\bibnamefont{Rubio}},
  \bibinfo{journal}{Phys.\ Rev. B} \textbf{\bibinfo{volume}{78}},
  \bibinfo{pages}{205425} (\bibinfo{year}{2008}).

\bibitem[{\citenamefont{Nakao}(1976)}]{nakao}
\bibinfo{author}{\bibfnamefont{K.}~\bibnamefont{Nakao}}, \bibinfo{journal}{J.\
  Phys.\ Soc.\ Japan} \textbf{\bibinfo{volume}{40}}, \bibinfo{pages}{761}
  (\bibinfo{year}{1976}).

\end{thebibliography}

\end{document}